\documentclass[acmsmall]{acmart}

\usepackage{siunitx}
\usepackage{breakurl}
\usepackage{subcaption}
\usepackage{algpseudocode}
\usepackage{graphicx}
\usepackage{tikz}
\usetikzlibrary{arrows,shapes,positioning}
\usepackage{multirow}
\usepackage{tabularx}
\usepackage{listings}
\usepackage{booktabs}
\usetikzlibrary{backgrounds}
\usepackage{textcomp}
\usepackage{amsmath}
\usepackage{algorithm}
\usepackage{algorithmicx}
\usepackage{algpseudocode}
\usepackage{hyperref}
\usepackage{makecell}
\usepackage{threeparttable} 
\usepackage{listings, xcolor}
\usepackage{enumitem}
\definecolor{verylightgray}{rgb}{.97,.97,.97}
\lstdefinelanguage{Solidity}{
  keywords=[1]{anonymous, assembly, assert, balance, break, call, callcode, case, catch, class, constant, continue, constructor, contract, debugger, default, delegatecall, delete, do, else, emit, event, experimental, export, external, false, finally, for, function, gas, if, implements, import, in, indexed, instanceof, interface, internal, is, length, library, log0, log1, log2, log3, log4, memory, modifier, new, payable, pragma, private, protected, public, pure, push, require, return, returns, revert, selfdestruct, send, solidity, storage, struct, suicide, super, switch, then, this, throw, transfer, true, try, typeof, using, value, view, while, with, addmod, ecrecover, keccak256, mulmod, ripemd160, sha256, sha3}, 
  keywordstyle=[1]\color{blue}\bfseries,
  keywords=[2]{address, bool, byte, bytes, bytes1, bytes2, bytes3, bytes4, bytes5, bytes6, bytes7, bytes8, bytes9, bytes10, bytes11, bytes12, bytes13, bytes14, bytes15, bytes16, bytes17, bytes18, bytes19, bytes20, bytes21, bytes22, bytes23, bytes24, bytes25, bytes26, bytes27, bytes28, bytes29, bytes30, bytes31, bytes32, enum, int, int8, int16, int24, int32, int40, int48, int56, int64, int72, int80, int88, int96, int104, int112, int120, int128, int136, int144, int152, int160, int168, int176, int184, int192, int200, int208, int216, int224, int232, int240, int248, int256, mapping, string, uint, uint8, uint16, uint24, uint32, uint40, uint48, uint56, uint64, uint72, uint80, uint88, uint96, uint104, uint112, uint120, uint128, uint136, uint144, uint152, uint160, uint168, uint176, uint184, uint192, uint200, uint208, uint216, uint224, uint232, uint240, uint248, uint256, var, void, ether, finney, szabo, wei, days, hours, minutes, seconds, weeks, years},  
  keywordstyle=[2]\color{teal}\bfseries,
  keywords=[3]{block, blockhash, coinbase, difficulty, gaslimit, number, timestamp, msg, data, gas, sender, sig, value, now, tx, gasprice, origin},  
  keywordstyle=[3]\color{violet}\bfseries,
  identifierstyle=\color{black},
  sensitive=false,
  comment=[l]{//},
  morecomment=[s]{/*}{*/},
  commentstyle=\color{gray}\ttfamily,
  stringstyle=\color{red}\ttfamily,
  morestring=[b]',
  morestring=[b]"
}
\usepackage{siunitx}
\newcommand\revision[1]{#1}

\newcommand{\name}{\textsc{LookAhead}}

\AtBeginDocument{%
  }

\setcopyright{cc}
\setcctype{by}
\acmDOI{10.1145/3729353}
\acmYear{2025}
\acmJournal{PACMSE}
\acmVolume{2}
\acmNumber{FSE}
\acmArticle{FSE083}
\acmMonth{7}
\received{2024-09-13}
\received[accepted]{2025-04-01}

\begin{document}

\title{LookAhead: Preventing DeFi Attacks via Unveiling Adversarial Contracts}

\author{Shoupeng Ren}
\orcid{0009-0000-6440-3800}
\affiliation{%
  \institution{Zhejiang University}
  \city{Hangzhou}
  \country{China}
}
\email{spren@zju.edu.cn}

\author{Lipeng He}
\orcid{0009-0002-1802-7394}
\affiliation{%
  \institution{University of Waterloo}
  \city{Waterloo}
  \country{Canada}
}
\email{lipeng.he@uwaterloo.ca}

\author{Tianyu Tu}
\orcid{0000-0002-9776-5610}
\affiliation{%
  \institution{Zhejiang University}
  \city{Hangzhou}
  \country{China}
}
\email{tianyutu@zju.edu.cn}

\author{Di Wu}
\orcid{0009-0007-4785-706X}
\affiliation{%
  \institution{Zhejiang University}
  \city{Hangzhou}
  \country{China}
}
\email{wu.di@zju.edu.cn}

\author{Jian Liu}
\orcid{0000-0001-6796-6828}
\thanks{*Jian Liu is the corresponding author.}
\authornotemark[1]
\affiliation{%
  \institution{Zhejiang University}
  \city{Hangzhou}
  \country{China}
}
\email{jian.liu@zju.edu.cn}

\author{Kui Ren}
\orcid{0000-0003-3441-6277}
\affiliation{%
  \institution{Zhejiang University}
  \city{Hangzhou}
  \country{China}
}
\email{kuiren@zju.edu.cn}

\author{Chun Chen}
\orcid{0000-0002-6198-7481}
\affiliation{%
  \institution{Zhejiang University}
  \city{Hangzhou}
  \country{China}
}
\email{chenc@zju.edu.cn}

\begin{abstract}

The exploitation of smart contract vulnerabilities in Decentralized Finance (DeFi) has resulted in financial losses exceeding $3$ billion US dollars. Existing defense mechanisms primarily focus on detecting and reacting to adversarial transactions executed by attackers that target victim contracts. However, with the emergence of private transaction pools where transactions are sent directly to miners without first appearing in public mempools, current detection tools face significant challenges in identifying attack activities effectively.

Based on the fact that most attack logic rely on deploying intermediate smart contracts as supporting components to the exploitation of victim contracts, novel detection methods have been proposed that focus on identifying these adversarial contracts instead of adversarial transactions. However, previous state-of-the-art approaches in this direction have failed to produce results satisfactory enough for real-world deployment.
In this paper, we propose \name, a new framework for detecting DeFi attacks via unveiling adversarial contracts. \name~leverages common attack patterns, code semantics and intrinsic characteristics found in adversarial smart contracts to train Machine Learning (ML)-based classifiers that can effectively distinguish adversarial contracts from benign ones and make timely predictions of different types of potential attacks.
Experiments on our labeled datasets show that \name~achieves an F1-score as high as $0.8966$, which represents an improvement of over $44.4\%$ compared to the previous state-of-the-art solution, with a False Positive Rate (FPR) at only $0.16\%$.

\end{abstract}

\ccsdesc[500]{Security and privacy~Intrusion detection systems}

\keywords{Blockchain, DeFi, Smart contract, Malware detection}



\maketitle

\section{Introduction}

{\em Decentralized Finance} (DeFi) has gained significant traction within the blockchain ecosystem over the recent years, stimulating the rise of a diverse array of DeFi applications such as: decentralized exchanges (DEXs)~\cite{UniswapV3, Balancer} and lending platforms~\cite{AAVE, Compound}.
The total value locked (TVL) in these DeFi protocols has exceeded \$100 billion in mid 2024~\cite{DefiLlama}. Among the various blockchains, Ethereum~\cite{wood2014ethereum} and Ethereum Virtual Machine (EVM)-compatible chains~\cite{Arbitrum, BNBChain, Optimism, Polygon} have formed a particularly thriving DeFi ecosystem.
At the heart of Decentralized Finance are {\em smart contracts}, which are self-executing programs deployed on blockchains, enabling trustless and transparent financial interactions without the need for intermediaries.
The functionalities of smart contracts are invoked by users through {\em transactions}, after which the blockchain consensus algorithms ensure predefined actions and state transitions are execute correctly and autonomously on-chain.
However, just like traditional software, smart contracts are fundamentally code-based programs, which are susceptible to vulnerabilities and various other security risks. In fact, DeFi incidents resulting from smart contract attacks have caused over 3 billion US dollars in financial losses~\cite{DeFiHackLabs, LearnEvmAttacks, zhou2023sok}.

\noindent
\textbf{DeFi Attacks.}
Targeting smart contracts, there are mainly two types of attack patterns commonly used by DeFi attackers: an attacker can either directly invoke specific public functions of the victim contracts in order to exploit a vulnerability~\cite{TreasureDAOHack, SushiSwapTokenHack}, or they can deploy an adversarial contract containing the complete attack logic onto the blockchain first, then initiate the attack through the intermediate contract by calling their entry functions afterwards~\cite{BeanstalkHack,BurgerSwapHack,PancakeBunnyHack,SushiswapExchangeHack}.

\noindent
\textbf{Detection Based on Adversarial Transactions.} Most existing real-time defense mechanisms leverage heuristics~\cite{qin2023towards,wu2023defiranger,zhang2020txspector,zhou2020ever} and Machine Learning (ML)~\cite{gai2023blockchain, wang2022defiscanner} techniques to detect and respond (e.g. via front-running) to yet-to-be-confirmed attack transactions in public mempools. The main issue with this transaction-based detection approach is that it faces significant challenges in applying to \emph{private adversarial transactions}.
Namely, attackers can utilize private mempool services to send transactions directly to miners, evading visibility from other participants of the blockchain network before it's confirmed and effectively bypassing attack detection and avoid any preventive actions from taking place.
Based on our empirical study of the historical DeFi incidents over the past few years, we found that out of the $142$ DeFi attacks on Ethereum collected by us, $80$ cases ($56.3\%$) involve transactions using private mempool services. Moreover, the proportion of adversarial transactions sent using private services is observed to exhibit a significant upward trend, rising from $44.4\%$ in the first half of 2022 to a noticeably higher $77.8\%$ in the latter half of 2023.

\noindent
\textbf{Detection Based on Adversarial Contracts.}
The ability for an attacker to execute private adversarial transactions severely undermines the detection capabilities of existing transaction-based solutions, rendering them largely ineffective. In contrast, some works have redirected their focus towards developing detection strategies based on identifying adversarial contracts. As an example, Forta~\cite{FortaNetwork} applies NLP techniques and uses a simple logistic regression model to analyze contract bytecode for malicious intents, but it fails to achieve satisfactory performance (Table \ref{tab:classifier_model_performance_comparison}). Beyond Forta, the only comparable work is BlockWatchDog~\cite{yang2024uncover}, which only support detecting reentrancy attacks. We emphasize the critical importance of designing an effective solution for detecting a wide range of DeFi attacks that can remain functional even when private mempool services are employed by the attackers.

\noindent
\textbf{Our method.} In this paper, we propose the \name~system, a new framework for detecting DeFi attacks via unveiling adversarial contracts. It achieves high detection effectiveness through systematic feature design and selection, the construction of a comprehensive contract dataset, and a combination of advanced classifier models. 
In \S\ref{thread model}, we describe in detail the threat model assumptions and the scope of detection followed by our proposed system.

We make an empirical observation that adversarial contracts used in DeFi attacks often contain similar patterns ($>70\%$ of the adversarial contracts use funds from anonymous sources for deployment; $>98\%$ are closed-sourced, etc.) We also note that depending on the vulnerabilities targeted, attackers write contract code in specific manners. 
For instance, flashloan attacks rely on complex call chains and the implementation of attack logic within callback functions.
To use these behaviours to our advantage for identifying adversarial contracts, we using the \emph{stacking} technique to design three ML models (transformer, classifier, and a meta classifier) that, when used together, can understand the intrinsic characteristics of malicious contracts based on code semantics and common patterns. To support the models in understanding the behaviour of Solidity code, we develop a specialized tokenization method called Pruned Semantic-Control Flow Tokenization (PSCFT) that is integrated directly into a contract bytecode decompiler.

An important observation made by~\cite{zhou2023sok} is that most attacks are not executed atomically within the \textit{constructor} during adversarial contract deployment, which provides a rescue time frame (after an adversarial contract has been deployed and before the attacker execute the attack logic via the contract) for defenders and victims. We use $tx_{deploy}$ to denote a contract deployment transaction, and $tx_{first}$ to represent the transaction in which the attacker initiates the actual attack via the adversarial contract. In our dataset, over $60\%$ of attacks exhibit a difference between the confirmation time of $tx_{deploy}$ and $tx_{first}$ that satisfies: $t_{first} - t_{deploy} \geq 100 s$. By designing an efficient feature extraction pipeline and leveraging modern ML architectures, we can produce predictions for newly created contracts on-chain in a timely manner with amortized $t^{pred} \ll 100 s$. The result of our design is \name, a system that is able to effectively distinguish adversarial contracts from benign ones, and make just-in-time predictions ($t_{deploy} + t^{pred} < t_{first}$) of potential attacks.

To ensure the performance and generality of our ML models, we hand-picked then carefully reviewed and identified $375$ adversarial contracts used in DeFi incidents between April 2020 and July 2024 from both the Ethereum and Binance Smart Chain (BSC) to form our adversarial contract dataset. And in total $\num{796,437}$ contracts deployed on Ethereum from June 2022 to June 2024 were collected, with $210,643$ of them being used as benign samples after filtration. We extract a selection of features from the contracts capturing patterns exhibited during both their implementation and their deployment stages. Given that our dataset contains a combination of contracts from both recent and historical attacks, we believe that our feature selection will remain reasonably relevant and effective even in the rapid growing landscape of decentralized finance.

We build our classifiers using multiple supervised ML algorithms and conduct a comprehensive performance evaluation. Experiment results on our labeled datasets show that our method for identifying adversarial contracts performs exceptionally well with an F1-Score produced by the KNN meta classifier reaching as high as $0.8966$, and a false positive rate of only $0.16\%$, which represents an improvement of over $44.4\%$ compared to the previous state-of-the-art solution Forta.

\noindent
\textbf{Contributions.} We summarize our contributions as follows:
\begin{itemize}
    \item We highlight the limitations in existing transaction and contract-based detection methods and propose a new framework \name~that achieves significantly better performance;
    \item We design a smart contract bytecode lifting and analysis pipeline with an integrated protocol to extract features and generate PSCFT for Natural Language Processing (NLP) training;
    \item We construct the first large-scale comprehensive dataset of adversarial (manually-labelled) and benign (methodically selected) smart contracts consisting of a set of useful contract features based on extensive empirical observations with statistical data support;
    \item We build and evaluate an ML-based system consisting of two open-weights classifiers and one meta classifier model, the results are compared with previous state-of-the-art works to substantiate our system's effectiveness and practicality.
\end{itemize}
\section{Background}
\label{background}

\subsection{Ethereum \& Decentralized Finance (DeFi)}

Blockchain is a decentralized ledger maintained over a peer-to-peer network via consensus mechanisms. Ethereum~\cite{wood2014ethereum} is one of the most used blockchain platforms. It was the first to introduce the Ethereum Virtual Machine (EVM) that support smart contracts, and has inspired the development of a series of EVM-compatible chains. DeFi is a blockchain-based financial ecosystem powered by smart contracts to offer financial services in a more open and transparent manner.

\textbf{Accounts.} Accounts are entities that can hold the Ether token and initiate transactions. Ethereum has two types of accounts: External Owned Accounts (EOAs) and Contract Accounts (CAs). EOAs are controlled by users holding private keys, while CAs are containers of contract code and storage.

\textbf{Transactions.} 
Transactions are used to transfer Ether, invoke contracts, or create contracts. After a transaction is sent, it is first placed in the \textit{mempool} of a blockchain node, waiting to be selected and finalized by block producers.

\textbf{Smart Contracts.} Smart contracts are instantiated objects stored on the blockchain. They are typically written in languages like Solidity~\cite{dannen2017introducing} and compiled into bytecode that can be executed within the EVM, a stack-based virtual machine supporting Turing-complete instructions.

\textbf{Tokens.}
Tokens are cryptocurrencies created using smart contracts. They come in two main types: fungible tokens and non-fungible tokens (NFT). Furthermore, Ether, used as the reward for anyone contributing in the consensus, is referred to as the native token.

\textbf{Decentralized Exchange (DEX).} 
Unlike centralized exchanges (CEX), DEX doesn’t require users to deposit digital assets into the exchange for trading. Instead, it conducts asset transfers directly on the blockchain via smart contracts, enabling transparency, security, and decentralization.

\textbf{Flashloan.}  
Flashloan is a unique form of lending that leverages the atomic nature of blockchain transactions to allow users to borrow and repay cryptocurrency within a single transaction. This mechanism lets users temporarily possess large amounts of tokens for a small fee. Despite of its convenience, it has been exploited in numerous DeFi attacks~\cite{cao2021flashot,qin2021attacking,wang2020towards}.

\subsection{Maximal Extractable Value (MEV)}
Traditionally, miners determine the order of transactions in a block based on gas prices. Since the mempool is publicly accessible, users have the ability to influence the placement of their transactions by adjusting the gas price, allowing for front-running~\cite{eskandari2020sok, torres2021frontrunner}. 
The profit gained from such manipulation, beyond standard rewards and fees, is known as Maximal Extractable Value~\cite{daian2020flash}.

\textbf{Front-running.} 
Front-running is a fundamental means used to extract MEV, where bots monitor the public mempool for target transactions and raise gas prices to execute their own first. It has also been used in transaction-based intrusion prevention systems~\cite{qin2023blockchain, zhang2023your} to front-run adversarial transactions, thereby preventing DeFi attacks.

\begin{figure}[t!]
    \centering
    \includegraphics[width=0.98\textwidth]{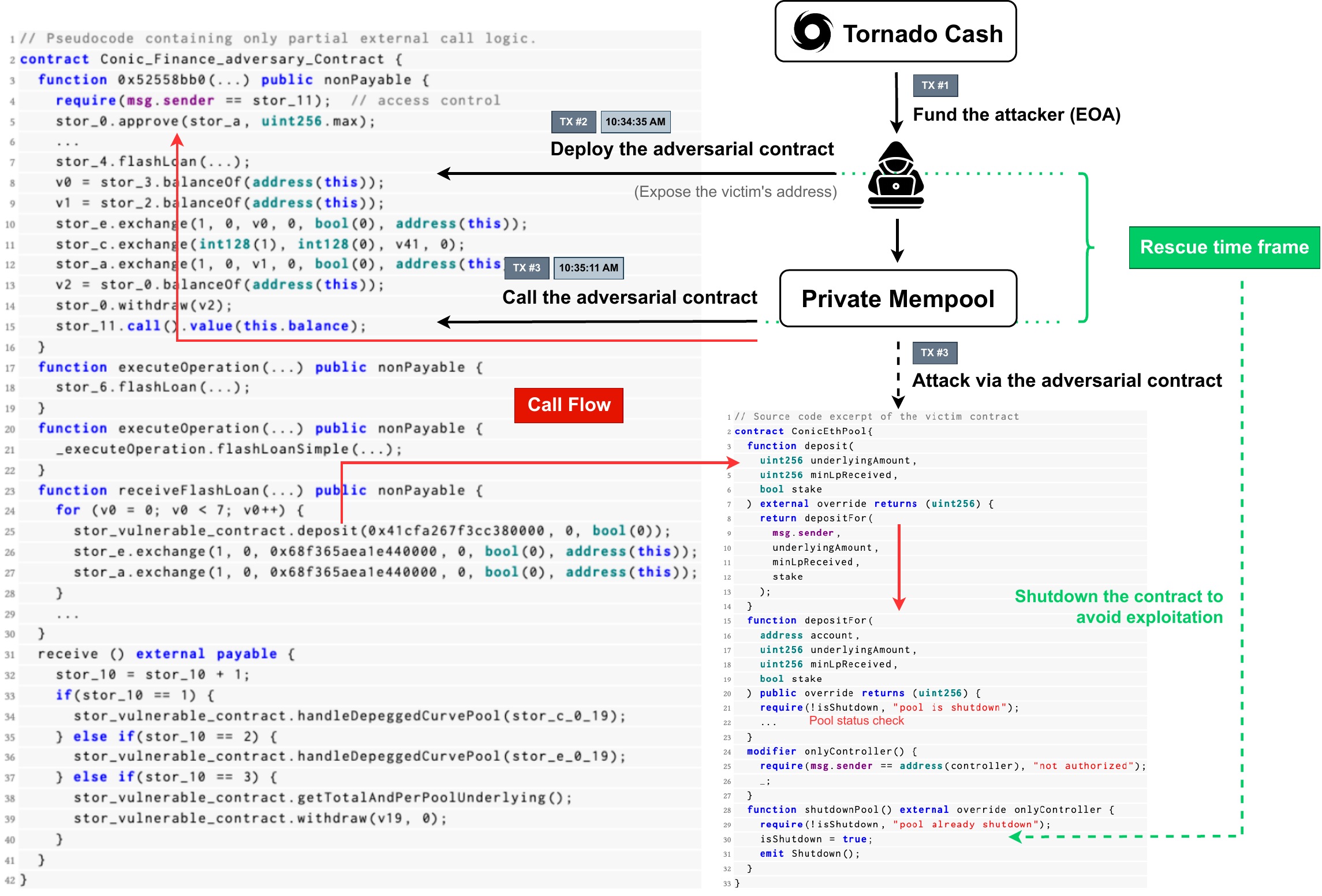}
    \caption{Read-only reentrancy attack on Conic Finance, using a private adversarial transaction to evade detection, could have been prevented during the rescue time frame.}
    \label{fig:example}
\end{figure}

\textbf{Private Mempool Services.} 
In response to MEV, Flashbots' MEV-auction has been widely adopted~\cite{Flashbots, weintraub2022flash}. This solution provides private channels for transactions to be submitted directly from the users to block producers without being broadcasted to the network, ensuring that transactions remain private until they are finalized, preventing them from being monitored or intercepted by others. 
However, these services have increasingly been abused by attackers to conceal adversarial transactions, effectively evading mainstream transaction-based detection methods.

\newcommand{\adv}{$\mathcal{A}$}
\section{Overview of \name}
\label{overview}

\subsection{Motivating Example}
On July 21, 2023, DeFi protocol Conic Finance suffered a major exploit~\cite{ConicFinanceHack} causing financial losses exceeding \$3 million. We present an overview of the attack process in Figure~\ref{fig:example}. Notably, the adversary transaction that triggered the attack was labeled as an \textit{MEV Transaction} on Etherscan~\footnote{\href{https://etherscan.io/tx/0x8b74995d1d61d3d7547575649136b8765acb22882960f0636941c44ec7bbe146}{https://etherscan.io/tx/0x8b74995d1d61d3d7547575649136b8765acb22882960f0636941c44ec7bbe146}}, indicating the use of private mempool services. This allowed the attacker to bypass the public mempool and invalidate any potential front-running defense attempts, highlighting the limitations of transaction-based prevention mechanisms that rely on detecting adversarial transactions.

\subsubsection{Rescue Possibility}
Despite the attacker using private mempool services to conceal the adversarial transaction, there was still an opportunity to prevent the attack, based on the following facts: 1) The ConicEthPool contract's address — the vulnerable target — was written in plaintext in the adversarial contract at deployment; 2) There was a 36-second window between the deployment of the adversarial contract and the execution of the adversarial transaction; 3) The ConicEthPool contract had a \textit{shutdownPool} function, which offers the ability to perform an emergency pause to block the adversary's attempt to call the \textit{deposit} function.
Thus, detecting the adversarial contract at the deployment stage, before the confirmation of the adversarial transaction, could have prevented the exploit, regardless of whether the attacker used a private mempool service or not.

\subsubsection{Adversarial Patterns}
While DeFi attack logic can vary based on the specific vulnerabilities being exploited, adversarial contracts often exhibit consistent behaviours due to the nature of their malicious intent and their profit motives.
Our analysis of numerous adversarial contracts reveals a clustered set of patterns including the use of anonymous fund sources, closed-source code, and frequent token-related function calls. Additionally, adversarial contracts often have distinctive interaction logic with victim contracts, which differs significantly from the patterns observed in normal contracts, resulting in easily distinguishable code structures and external call flows. Based on these observations, ML techniques can be employed to effectively identify adversarial contracts (regardless of the specific attack type) by learning and predicting these common patterns.

\subsection{Threat Model}
\label{thread model}

\subsubsection{DeFi Attacks}
\label{defi attacks}

There are many types of security breaches with complicated root causes, potentially leading to attacks on victims. 
In various scenarios, an adversary could use distinct techniques for exploitation. In this section, we present these attack methods, define key concepts involved in the attack process, and restrict \name's detection scope to a certain area of focus.

\begin{figure}[t!]
    \centering
    \includegraphics[width=0.7\textwidth]{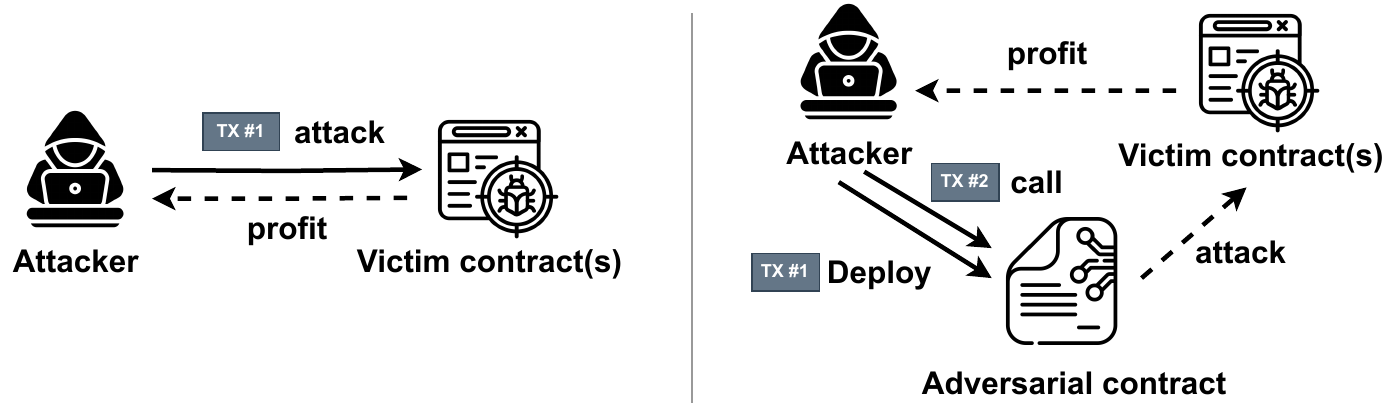}
    \caption{a) Left: attack directly, b) Right: attack through adversarial contracts.}
    \label{fig:DeFiAttack}
\end{figure}

\noindent
\textbf{Attack Methods.} The attack process of two common attack methods are depicted in Figure~\ref{fig:DeFiAttack}.

\renewcommand{\labelenumi}{\arabic{enumi})}
\renewcommand{\labelenumii}{\alph{enumii})}
\begin{enumerate} 
    \item \textbf{Attacking directly via \emph{adversarial transactions}.} In the presence of complex attacking logic involving multiple transactions, the attackers often bear the risk of encountering potential \emph{half-way failure}. 
    A half-way failure arises when only part of an attack was executed successfully and the rest fail to proceed due to various reasons (e.g., transaction conflicts).
    \item \textbf{Attacking through \emph{adversarial contracts}.} This pattern allows the attacker to execute complex attacking logic in a single transaction, hence avoid half-way failures thanks to the atomic nature of a blockchain transaction. Using this method, an attacker first deploys a smart contract on-chain containing the core attack logic, then initiate the adversarial transaction by calling the entry point function of the adversarial contract. Most DeFi attacks follow this pattern, hence we focus on addressing this type of attacks.
\end{enumerate}

\noindent
\textbf{Scope of Detection.} We specify two attack scenarios that fall outside of the scope this study. 1) Attacks that do not require contract deployments. Those resulting from private key leakage or attackers directly invoking public functions of the victim contracts using EOAs are outside the scope of our detection. 2) A small portion of attacks are completed directly through the \textit{constructor} function at the time of contract deployment. Since the adversarial contracts used in these attacks already exhibit execution traces of the actual attack at the time of deployment, we consider them as adversarial transactions, which fall outside the scope of our detection. 

\name~focuses on unveiling attacks that intend to cause financial loss for DeFi protocols by exploiting vulnerabilities in on-chain smart contracts. To identify an attack, our solution detects adversarial contracts that contain the core attack logic. Other contracts, such as those used solely for concealing profits, fake token contracts that facilitate the main attack, and counterattack contracts used by front-running bots will not be included for our analysis.

\subsubsection{Adversaries}

We consider a resource bounded adversary \adv~that is financially rational, capable of deploying an arbitrary amount of smart contracts and execute both public and private transactions on EVM-compatible blockchains.

\noindent
\textbf{Attacker Assumptions.} Based on previous research findings~\cite{zhou2023sok} (IEEE S\&P'23), our study makes the following assumption about \adv: 
\begin{enumerate}
    \item \adv~does not initiate attacks by batching $tx_{deploy}$ and $tx_{first}$ together in an atomic transaction during contract deployment, meaning that \adv~splits an attack into two stages following the second pattern described in Figure \ref{fig:DeFiAttack};
\end{enumerate}

In a prior research~\cite{parhizkari2023timely}, it was observed that 79.49\% of attacks in their dataset exposed the victim contract address before the adversarial transaction. Using the same method, we found 202 attacks (53.9\%) with similar exposure behaviour in our dataset of $375$ adversarial contracts. Hence, we also consider a weaker adversary $\mathcal{A}'$ satisfying one more assumption than \adv:

\begin{enumerate}
    \setcounter{enumi}{1}
    \item $\mathcal{A}'$ hard-codes exploitation targets' addresses in their adversarial contracts whenever possible, and the victim's address will emerge before the first adversarial transaction $tx_{first}$.
\end{enumerate}

\subsubsection{Victim of Attacks}

We consider those contracts which contain vulnerabilities that have been exploited by attackers and suffered a financial loss the victims of a \emph{DeFi attack}.

\subsubsection{Rescue Time Frame}
\label{Rescue Time Frame}

Our threat model leaves a short rescue time frame for detection and prevention to take place by the defenders and the targets, and enables the possibility of on-time reaction by victims of ongoing attacks. The rescue window is defined by $t^{window} = t_{first} - t_{deploy} - t^{pred}$, it has a typical value observed in our dataset of $t^{window} \leq 100$. Within the rescue time frame several counterattack actions could be taken, we describe those in more details in \S\ref{defense_mechanisms}.

\subsection{System Model}

\begin{figure}[t!]
    \centering
    \includegraphics[width=0.95\linewidth]{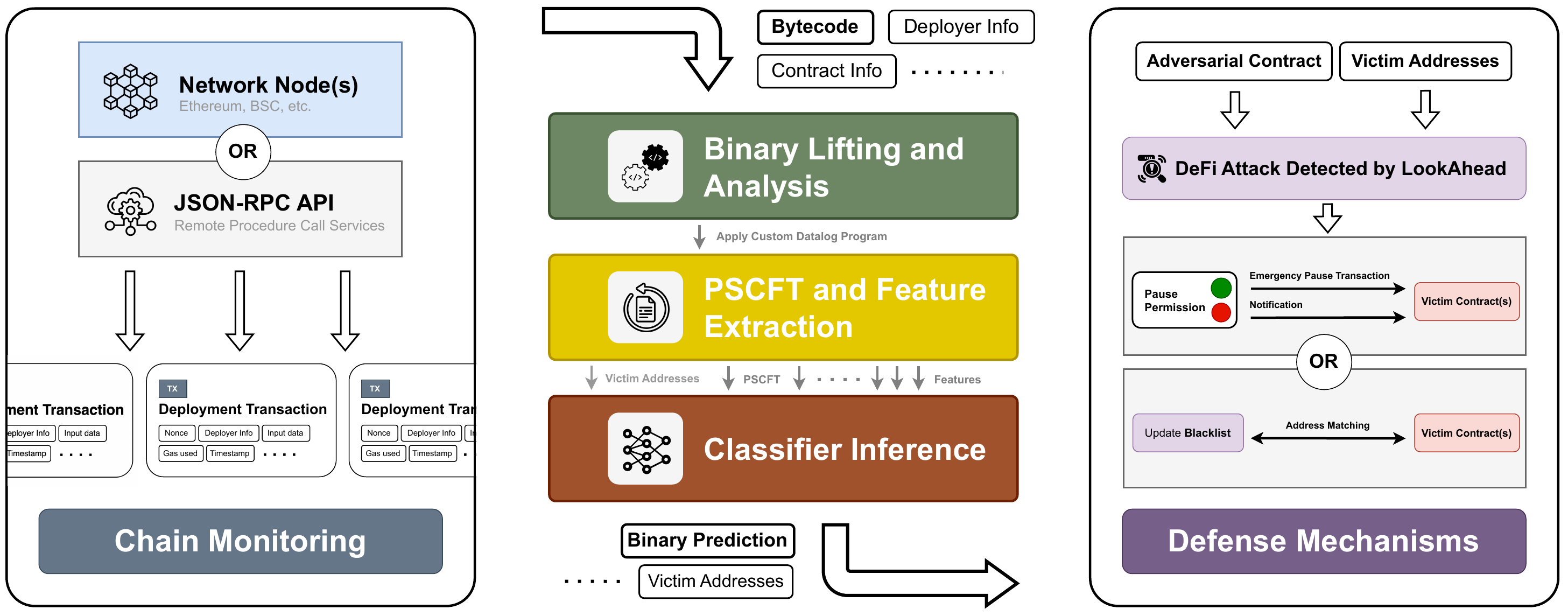}
    \caption{An overview of the \name~system.}
    \label{fig:system}
\end{figure}

In order for the \name~system to provide just-in-time identification of adversarial contracts as they are deployed onto the blockchain, a streamlined process consisting of a series of components is used. We show an overview of our system in Figure~\ref{fig:system}.

\textbf{Chain Monitoring.} Through a blockchain network node or third-party node monitoring services, we monitor on-chain data in real-time, and acquire the deployed contracts and their basic information such as deployer, input data, gas used, bytecode and so on whenever needed. Via chain explorers such as Etherscan, we are able to retrieve the verification status of contracts as well. These data will be combined and fed into a classifier as features for ML inference at a later stage.

\textbf{Binary Lifting.} We do not use any source code information from the contracts, but only the bytecode deployed on the blockchain to ensure our model is capable of analyzing any smart contract on EVM blockchains. To do that, \name~uses the binary lifting and analysis library \emph{Gigahorse}~\cite{grech2019gigahorse} to decompile the contract bytecode and generate intermeidate representation (IR). 

\textbf{PSCFT and Feature Extraction.} With the raw IR, we use a specialized feature extraction program to pull out features including but not limited to: token-related call information, flashloan usage information, etc. Finally, we convert the IR into a condensed text form (PSCFT) and populate it with semantic information such as function names and external call target address labels.

\textbf{Classifier Inference.} Input features are properly tokenized and formatted before they run through our trained classifiers. These ML models are made up of a transformer, an optimal candidate classifier, and a meta classifier trained using supervised machine learning methods for binary classification. They are able to output either the classification result $label_{pred} \in \{0, 1\}$ indicating whether the contract is adversarial ($1$) or not ($0$), or they can produce a confidence score $0\% \leq p_{pred} \leq 100\%$ representing the possibility of the input contract being adversarial.

\textbf{Defense Mechanisms.}\label{defense_mechanisms} After detecting an ongoing attack, \name~enables several potential defense mechanisms for defenders and victim contract owners: 1) To defend against $\mathcal{A}'$, notifications are sent to targets, allowing them to take defensive actions, such as executing an emergency pause. If contract owners grant admin permissions to \name, the system can autonomously trigger shutdown methods on behalf of them. 2) To mitigate against \adv, we propose a blacklist service for DeFi protocols, enabling them to block transactions from flagged adversarial contracts.
\section{Dataset Building}
\label{dataset}

\subsection{Contract Collection}
\label{Contract Collection}

To perform supervised training of our classifier models, we collect a large set of smart contracts from the Ethereum and BSC. For the positive (adversarial) dataset, we curated a collection of adversarial contracts used in real-world DeFi incidents. For the negative (benign) dataset, we follow a methodical process to collect contract samples that are very unlikely to be related to any attacks.

\subsubsection{Adversarial Contracts}
\label{Adversarial Contracts}
To collect adversarial contracts, we gather information on DeFi incidents from various publicly available sources, such as GitHub repositories documenting DeFi incidents \cite{DeFiHackLabs, LearnEvmAttacks}, and alerts from security companies \cite{blocksec, certik, slowmist}. We conduct a meticulous review of these incidents to identify the contracts used by attackers.
Between April 2020 and July 2024, we collected 506 \textit{DeFi attacks}. In these, $29$ attacks ($5.7\%$) involved direct attacks from EOAs without using contracts, and $37$ attacks ($7.3\%$) were executed directly within the contract \textit{constructor}. Both cases fall outside the scope of our detection. While attacks spanned multiple blockchains, the majority were on Ethereum and BSC. To simplify the study, we exclude 65 incidents happened on other chains. This left us with $375$ adversarial contracts: $142$ from Ethereum and $233$ from BSC.
Following prior research~\cite{zhou2023sok}, we classify the attacks into three categories: 1) Untrusted or unsafe calls ($13.3\%$), 2) Access control mistakes ($14.4\%$), and 3) Coding mistake ($72.3\%$). The first category includes issues like reentrancy, the second covers access control flaws, and the third involves coding mistakes such as arithmetic errors and logic absences.

\subsubsection{Benign Contracts}
\label{Benign Contract Collection}

As contract-based \textit{DeFi attack} detection approach is still a largely under-explored research direction, to date, there is no publicly available dataset that provides a comprehensive and representative set of ground truth benign contracts. 

Consequently, we need to build our benign dataset from scratch. Given the vast number and variety of smart contracts deployed on the blockchain, constructing a large-scale 
dataset with sufficient sample diversity and label accuracy requires extensive manual review efforts and is prohibitively time-consuming. 
In response to this, previous ML-based
work Forta \cite{FortaNetwork} follow a heuristic approach for collecting benign samples.

\noindent \textbf{Previous Approach.} Forta assumes that all contracts verified on the chain explorer are benign. Despite being an intuitive design choice, analysis of our attack samples reveals that a small proportion of adversarial contracts are nonetheless verified. 
This may be because verification is an inexpensive task (often automated by development toolkits) and it has negligible impact on attack effectiveness, as existing detection methods do not focus on identifying adversarial contracts. 
Consequently, a more robust collection method is needed.

\noindent \textbf{Key Observations.} A smart contract can be deployed either 1) through a normal transaction, where its \textit{to\_address} field is \textit{NULL} with bytecode included in the \textit{input\_data} field, or 2) via an internal transaction (also known as a message call) initiated by an existing contract during its execution. Between June 2022 and June 2024, the number of contracts created via internal transactions is significantly higher than those created via normal transactions (approximately 20 times more).
A prior research~\cite{gupta2019study} that analyzed $1.16$ million contracts created via internal transactions found that these contracts were generated by only $9,228$ creator contracts, indicating that a small number of contracts were responsible for deploying a vast number of contract instances. Most of these instances originate from factory contracts, e.g., Uniswap’s factory contract~\footnote{\href{https://etherscan.io/address/0x5c69bee701ef814a2b6a3edd4b1652cb9cc5aa6f}{https://etherscan.io/address/0x5c69bee701ef814a2b6a3edd4b1652cb9cc5aa6f}}, which are designed to repeatedly create similar contracts. Additionally, we observe that all adversarial samples in our dataset are deployed through normal transactions. These contracts are typically invoked by a very limited number of unique addresses. Specifically, $355$ of them ($94.7\%$) interacting with only a single address, namely the attacker's account, and a maximum of $6$ unique users. This low interaction count is consistent with the behavior of adversarial contracts, which typically feature access control and are designed for single-use only to attack specific victims.

\noindent \textbf{Our Method.} 
Based on the observations, we construct our benign dataset following two key principles to ensure evaluation fairness, as well as dataset diversity and accuracy.

\begin{itemize}[left=17pt]
    \item To ensure fair evaluation, we focus solely on contracts created via normal transactions, consistent with the deployment patterns of all adversarial samples, while avoiding internal transactions that could drastically inflate the dataset with numerous near-identical instances.
    \item We select contracts with over a defined number of unique interacting addresses as an indicator of their benign nature, aiming to ensure reliability while maintaining sample diversity.
\end{itemize}

\noindent
We first query Google's BigQuery service to collect contracts deployed via normal transactions between June 2022 and June 2024, identifying $796,437$ contracts. Then we analyzed their transaction history, focusing on the number of unique interacting addresses from June 2022 to August 2024, and retained contracts with at least $10$ unique interacting addresses, resulting in a final benign dataset of $210,643$ contracts.
While no heuristic method can fully eliminate potential bias, our approach offers a pragmatic trade-off between dataset scale, diversity, and correctness. 

\subsection{Feature Selection}
\label{Feature Selection}

Based on empirical observations, we construct a selection of features for use in \name's ML model training and inference. The set consists of features spanning both the implementation and the deployment stages of a \emph{DeFi attack}, aiming to capture any potential unexpected patterns of behaviour exhibited by attackers. 
In this section, we explain our rationale behind the choice of features and provide supporting statistical data whenever applicable.

\subsubsection{Deployment Features}
\label{DeploymentFeatures}

A typical workflow for deploying a contract to a blockchain consists of 1) Sourcing the fund to cover deployment costs, 2) Sending a transaction containing the contract bytecode and other related data on-chain, 3) Verifying the source code of the contract to demonstrate its security and trustworthiness to the wider community. During the process, a malicious individual often behave very differently from a normal one, we use this to our advantage by selecting several measurable properties as supporting factors to help us make such a distinction.

\textbf{Creator Attributes:} $\{nonce, fund\_source\}$. $nonce$ indicates number of transactions sent from the contract creator before the deployment of the contract. A higher $nonce$ suggest a more active EOA address. We are also interested in the $source$ from which the contract creator initially obtained their funds. To conceal their identities, attackers typically fund their attacks from a mixer (e.g, Tornado Cash~\cite{wang2023zero}) or an instant exchange~\cite{ChangeNOW, SideShift} without sign-up or KYC (Know Your Customer) requirements, then transfer the funds to a new account for subsequent transactions. In contrast, regular users are more likely to source their funds from KYC-required centralized exchanges.

We categorize fund sources into four types based on address information:

\begin{enumerate}
    \item \textbf{Safe:} Fund sources with relatively secure origins, such as KYC-required centralized exchanges, which are less likely to be used by attackers.
    \item \textbf{Anonymous:} Fund sources with anonymous origins, including mixers and some KYC-free exchanges, which are more susceptible to be used for malicious purposes.
    \item \textbf{Bridge:} Funds originating from cross-chain transfers. The source could be anonymous, hence they are considered potentially usable for attacks.
    \item \textbf{Unknown:} Fallback label for those that fail to fall into the above categories.
\end{enumerate}

\textit{Statistical Data:}
In benign samples, we observe that more than $75\%$ of them have Safe sources, and approximately $10\%$ of them are funded through Anonymous sources. In contrast, more than $70\%$ of the adversarial contracts have Anonymous sources, with only about 10\% having Safe labels.

\textbf{Transaction Data:} $\{value, input\_data\_length, gas\_used\}$. Depending on the contract's design, the contract creator can transfer native tokens to the contract upon deployment. Adversaries typically do not perform such a transfer. To avoid the potential discrepancy across Ethereum and BSC, we binarize the \textit{value} attribute to indicate whether such a transfer occurred (true or false), rather than using the exact amount.
The input data field of a deployment transaction holds the contract's bytecode, we measure its length to account for the size of the contract. Additionally, we use $gas\_used$ to measure the computational resources consumed during deployment, which depends on the bytecode complexity and EVM execution, rather than gas fees, which relies not only on $gas\_used$ but also on the dynamically fluctuating gas price influenced by network congestion. 

\textit{Statistical Data:} In our dataset, about 80\% of benign contracts have input data exceeding 15,000 bytes, while over 50\% of adversarial contracts fall below this threshold. To eliminate potential EVM-induced differences in $gas\_used$ between Ethereum and BSC, we simulate the deployment of all BSC adversarial samples on Ethereum using identical bytecode to obtain a consistent \textit{gas\_used} value. However, 24\% of the contracts failed due to interactions with contracts absent on Ethereum. Among the successfully samples, 94\% showed less than a 1\% deviation compared to original value. Given this negligible difference, for the failed samples, we directly use their original \textit{gas\_used} value.

\textbf{Verification Status:} $\{verified\}$. The $verified$ status can be granted to contracts if their source code and compiler settings are submitted to platforms like Etherscan. Verified contracts are open-source, anyone can view their code. Many contract development frameworks, such as Foundry~\cite{foundry}, offer options to verify contracts upon deployment. It has become increasingly common for developers to verify their contracts to demonstrate security and trustworthiness. Attackers, on the other hand, typically do not have the incentives to do so since it would expose their malicious intentions.

\revision{\textit{Statistical Data:} We observe in our dataset that only $6$ adversarial contracts ($1.6\%$) are verified , while more than $85\%$ of benign contracts are verified.}

\subsubsection{Implementation Features}
\label{Implementation Features}

Based on empirical observations, some of the characteristics that can be used to determine whether a contract is adversarial or not include: implemented functions, internal and external function calls. To capture other hidden or intrinsic properties of a contract, we also design a special technique called Pruned Semantic-Control Flow Tokenization (PSCFT) that provides a condensed representation of the behaviour of smart contracts.

\textbf{Implemented Functions:} $\{func\_count, flashloan\_callback\_count\}$. Since adversarial contracts typically do not involve complex business logic, they often implement fewer functions. We also calculate the number of flashloan callback functions and the proportion they take up over all public functions. This choice is motivated by the fact that many attack scenarios require attackers to have significant token holdings, which most adversaries do not naturally possess. Consequently, they resort to utilizing flashloan to temporarily borrow a substantial amount of tokens and execute attack logic within corresponding callback functions. Since the size of lending pools are limited, attackers may borrow tokens from multiple sources, resulting in multiple flashloan callbacks.

\revision{\textit{Statistical Data:} 
We can observe from the dataset that over 75\% of benign contracts incorporate at least 20 public functions, whereas more than 75\% of adversarial contracts implement 10 or fewer. Additionally, over 60\% of adversarial contracts include flashloan callbacks, while more than 99\% benign contracts involve no such callbacks.}

\textbf{Function Calls:} $\{token\_call\_count, max\_token\_call\_count, avg\_token\_call\_count, delegate\_c\\all\_count, selfdestruct\_count\}$. We take into consideration external calls related to tokens, including their count and proportion. Generally, the objective of DeFi attacks is to gain financial benefits, i.e., acquire valuable tokens. Therefore, external calls within adversarial contracts often involve token-related functions. The token-related functions we selected include, but are not limited to, common token standard~\cite{ERC20, ERC721} functions and popular DEX~\cite{UniswapV3} functions.

Furthermore, we introduce two additional features to assess the behaviour of public functions in a smart contract. $max\_token\_call\_count$ is designed to capture the maximum number of token-related interactions within the execution flow of any public function of a contract, which helps us identify contracts with exceptionally high token interactions. Moreover, $avg\_token\_call\_count$ is also measured to demonstrate the average number of token-related calls per public function.

We also include the number of delegatecall instructions, as these are often implemented in proxy contracts. Additionally, some attackers have a tendency to self-destruct their contracts after completing the attack, so we also consider the presence of the selfdestruct instruction.

\textit{Statistical Data:} It can be observed from our dataset that over $90\%$ of benign contracts do not have token-related calls. In contrast, almost all adversarial contracts involve token-related calls, with over $70\%$ of them having $10$ or more.

\begin{figure}[t!]
    \begin{subfigure}{0.385\linewidth}
        \centering
        \includegraphics[width=1\linewidth]{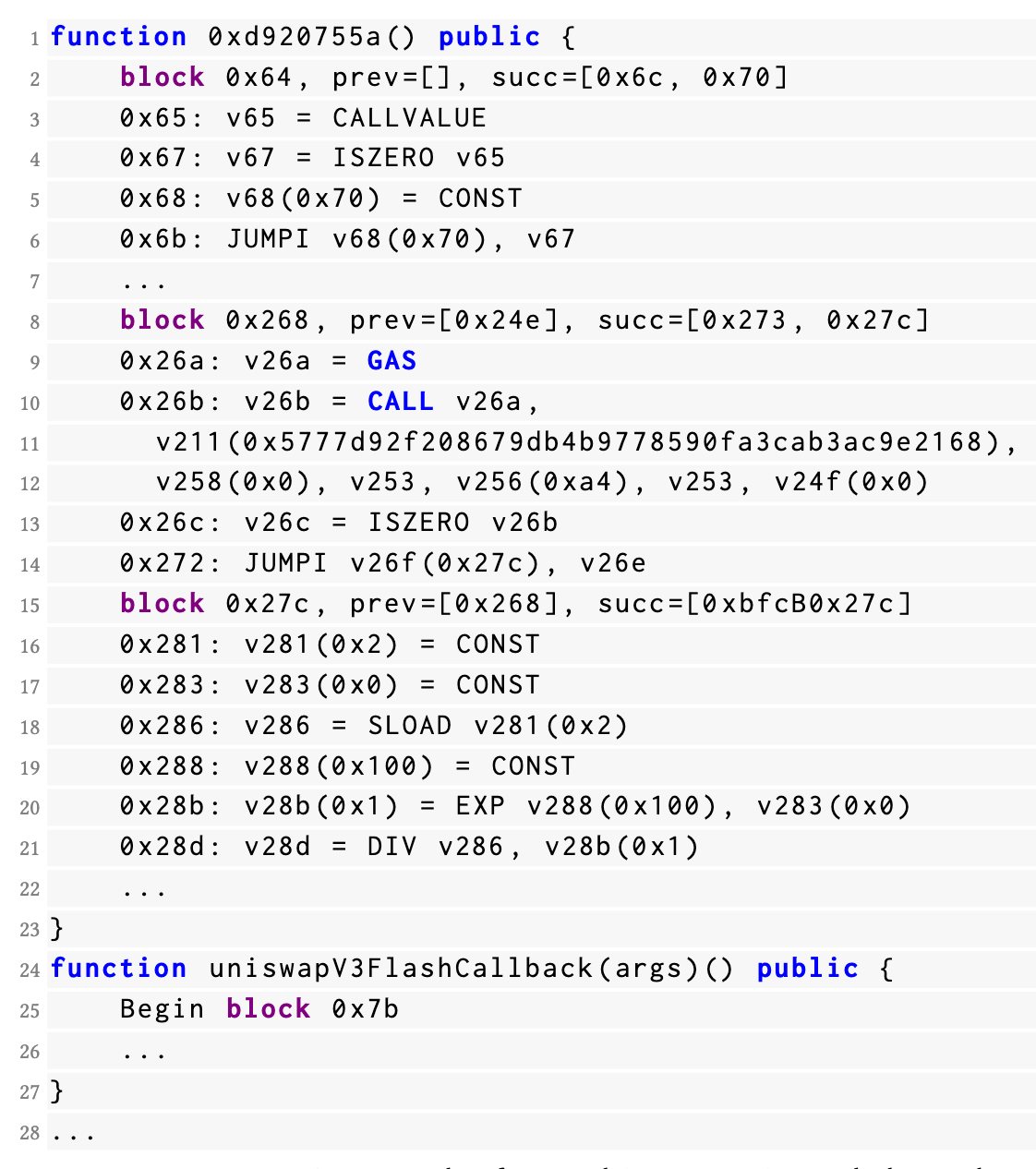}
    \end{subfigure}
    \hspace{0.05\linewidth} 
    \begin{subfigure}{0.34\linewidth}
        \centering
        \includegraphics[width=1\linewidth]{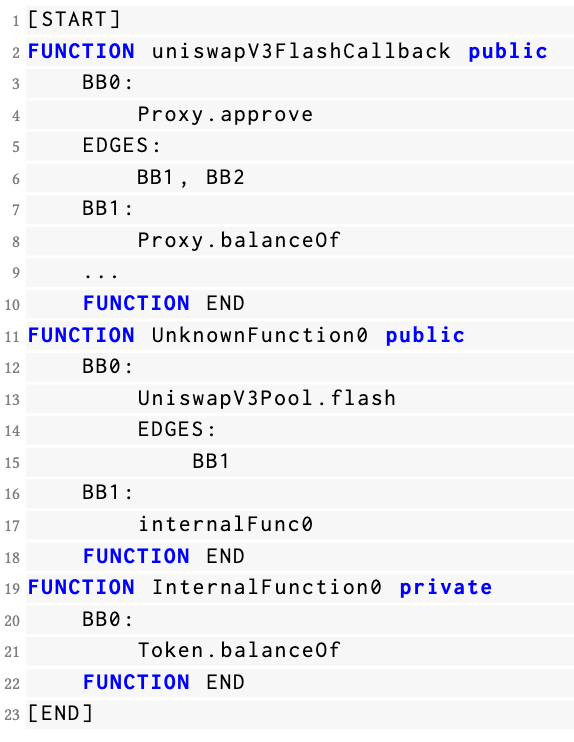}
    \end{subfigure}
    \caption{Side-by-side comparison of raw IR and PSCFT.}
    \label{fig:ir_pscft_comparison}
\end{figure}

\textbf{Code Semantics:} $\{PSCFT\}$. To understand the behaviours exhibited by \emph{DeFi attack} adversaries and capture features that are potentially hidden from our empirical observations, we design a new representation for smart contracts called Pruned Semantic-Control Flow Tokenization ($PSCFT$).

In smart contracts, functionalities (including potential attack logic) are implemented through \textit{functions}, and interactions with external entities, such as invoking the functions implemented in other deployed contracts, are performed via \textit{message calls}. For adversarial contracts, such interactions (e.g., with victim contracts) are indispensable. 
Therefore, each function, along with its message calls and control flows, is capable of capturing the intrinsic behavioral characteristics of a contract, providing key information for determining whether the contract is adversarial or benign.

To extract these critical features, we use Gigahorse \cite{grech2019gigahorse} to lift the contract bytecode into an intermediate representation (IR), followed by a pruning process that remove extraneous elements while reconstructing control flows. To provide richer contextual information, the refined IR is fed into an augmentation pipeline that retrieves and replaces function signatures and external addresses with appropriate semantic labels. The resulting output is the fully PSCFT, which effectively represents a textual summary of function canonical names, control-flow graphs and external call chains. Figure~\ref{fig:ir_pscft_comparison} presents an excerpt comparing the raw IR of a contract with the corresponding PSCFT. We provide the detailed construction process of PSCFT in \S\ref{PSCFT method}.
\section{Methodology}
\label{methodology}

\begin{figure}[t!]
    \centering
    \includegraphics[width=0.95\textwidth]{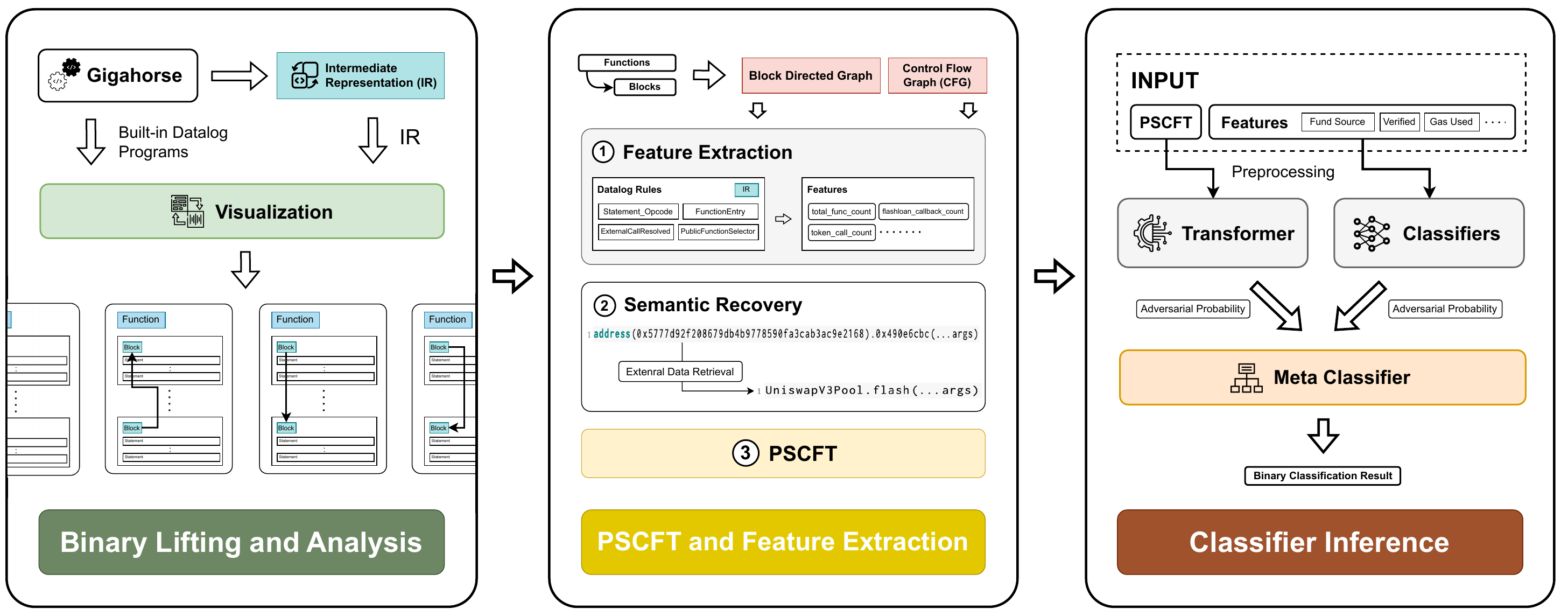}
    \caption{\name's system building workflow}
    \label{fig:workflow}
\end{figure}

In this section, we present our methodology for building the \name~system that performs adversarial contract classification. An overview of the complete framework is depicted in Figure~\ref{fig:workflow}.

\subsection{External Data Retrieval}
\label{External Data Retrieval}

Some of the features used in our dataset are obtained from external sources, we identify related third-party services in this section. Based the data, we use a algorithm to determine fund source for the contracts, enabling subsequent training and inference of our machine learning models.

\textbf{Transaction Data.} In this study, we focus on investigating and demonstrating the viability and effectiveness of the \name~system. Therefore, for the sake of simplicity, instead of running our own blockchain nodes, we obtain transaction raw data from an RPC (Remote Procedure Call) provider such as Alchemy~\cite{Alchemy} that relays our requests directly to other nodes in the blockchain network. Prior to the practical deployment of \name, we recommend setting up dedicated nodes for minimized data processing delays. We store and format contract deployment transactions and their associated attributes such as nonce, value, and gas used, along with the sender and input data properties for use in subsequent phases of our feature extraction pipeline.

\textbf{Chain Explorer Data.} We leverage the APIs provided by chain explorers, Etherscan and BscScan to retrieve information about the verification status of contracts. However, we cannot obtain the exact time for contract verification. The effectiveness of this feature is based on the assumption that developers verify their contracts upon deployment (with minimal to no delay). We describe the potential impact that this assumption could bring to the performance of \name~in \S\ref{effectiveness}.

Furthermore, we develop a fund source labeling algorithm to determine a contract's fund source. 
We first construct an address label dataset using the address label cloud~\cite{labelcloudeth, labelcloudbsc} and the public name tags provided by chain explorers, following the definitions in~\S\ref{DeploymentFeatures}. As an example, Tornado.Cash~\cite{Tornado_cash} is given the label of Anonymous. The algorithm traces fund flows from the contract deployer, recursively identifying the earliest funding source. If the source address has a known label, it is assigned; otherwise, tracing continues up to a predefined threshold depth. If no label is identified within this limit, the fund source of the contract is marked as Unknown.

\subsection{Pruned Semantic-Control Flow Tokenization (PSCFT)}
\label{PSCFT method}

Based on EVM bytecode binary lifter framework Gigahorse~\cite{grech2019gigahorse}, we design a pipeline for 1) Constructing intra-contract Control Flow Graphs (CFG) and external call chains based on Intermediate Representation (IR), 2) Aggregating and extracting dataset features from EVM bytecode, and 3) Generating textual summaries for the smart contracts in our dataset.

\subsubsection{Implementation Feature Extraction}
\label{Implementation Feature Extraction}

To enable the extraction of implementation features (see \S\ref{Implementation Features}) simultaneously with the generation of the IR, we implemented a custom-designed Souffle Datalog program based on existing rules from Gigahorse. 1) First, by utilizing the \textit{Statement\_Opcode} rule, which maps each statement to its corresponding opcode, we can directly extract the number of occurrences of critical opcodes, such as \textit{DELEGATECALL} and \textit{SELFDESTRUCT}, which corresponds to features $delegate\_call\_count$ and \textit{selfdestruct\_count}. 2) Next, we leverage the \textit{FunctionEntry} rule, which contains all function entry points (identifiers), with the help of \textit{PublicFunctionSelector}, which maps public functions to their selectors, to determine the total number of public and private functions. This combination also helps identify flashloan callback functions. 3) Finally, to analyze the external function calls, we design the \textit{ExternalCallResolved} rule, providing details about all resolved external calls made within the contract, including their selectors and signatures. 
With the help of \textit{ExternalCallResolved}, we can keep track of token-related calls and their related information including their total count, maximum occurrences, and average usage.

\subsubsection{CFG Construction and Pruning}

Based on the IR of a contract, we construct a function-by-function visualization $\mathcal{F} = \{\mathcal{F}^{private} \cup \mathcal{F}^{public}\}$ of the internal structure of contract methods, each consisted of control flow \textit{blocks} and programming \textit{statements}. Functions have 4-byte selector derived from the Keccak hash (SHA3) of their canonical names and can be either private or public. Each function contains multiple basic blocks (BB) chained together ($\{F_i = (BB^i_0, BB^i_1, \cdots, BB^i_{|F_i|})\}$), and each block contains multiple statements. Blocks and statements are identifiable via unique numerical IDs, for example, $0x11b9$. The relationship between two blocks can be discovered by following the predecessor and successor IDs associated with each block, each block can have multiple predecessors and successors, hence we can create a directed graph $G_i = (B_i, A_i)$ such that the set of nodes $B_i = \{BB^i_j : BB^i_j \in F_i\}$ where $F_i \in \mathcal{F}$, and $(BB^i_x, BB^i_y)$ is an arc in $A_i$ if $BB^i_x$ is a predecessor of $BB^i_y$ and $BB^i_y$ is a successor of $BB^i_x$. Statements could be constant/variable declarations, internal/external function calls, function terminations, memory copies, storage initialization, etc., we focus on constructing a CFG based on function call statements.

The pruning process starts with the sorting and reordering of functions based on function types (public first, then private) and their canonical names. Gigahorse can only recover the names for some of the public functions and external call methods, hence if the derivation of canonical names from function selectors failed, we will leave them as-is and recover their names in the semantic recovery stage. For any private function $F^{private}_j \in \mathcal{F}^{private}$, we give it a canonical name of $InternalFunction_i$. Next, we remove all statements not related to function calls from every block $BB^i_j \in B_i$. Then, we perform a depth-first search (DFS) over $G$ to remove any block $BB^i_k$ that contains no statements (i.e. $|BB^i_k| = 0$) from $B_i$, along with its incoming and outgoing arcs from $A_i$, and update the the predecessor and successor IDs for its neighbours. Finally, we assign canonical names $BB^i_j, j \in [0, |B| - 1]$ to basic blocks that are still left in $B_i$.

\subsubsection{Semantic Recovery}

For an external call statement like the one shown in \S\textit{PSCFT and Feature Construction} in Figure \ref{fig:workflow}, we aim to recover: 1) An address label for \textit{0x5777d92f208679db4b9778590\\fa3cab3ac9e2168}, 2) A human-readable name for the target method \textit{0x490e6cbc}.


Similar to our fund source labelling algorithm, we consider to match the target address with an address associated with a known label in chain explorers' address label cloud and public name tags. For contracts that cannot be matched with an existing label, we check if they are open source, when they are, we use their contract name as the label. Contract developers might choose to store the target address in the EVM storage or as a variable/constant,
in which case we use the same technique as~\cite{kong2023defitainter} to run a Datalog program that finds the slot location storing the address value and retrieves the value from the target storage. If this process did not result in a matching label, we assign the address a default label \textit{UnknownTarget}.

Function selectors are generated by hashing the function signatures with a secure hashing algorithm, and the hashing process is a one-way trap door. However, thanks to the deterministic nature of hash functions, third-party service such 4byte~\cite{4bytes} are able to construct and maintain a large database of publicly known hash-to-name mappings. Therefore, in hopes to include as much function names and as little unknown functions in the PSCFT as possible, we attempt to extend Gigahorse's name recovery ability with 4byte's service by sending a request to their API upon failed recovery by Gigahorse. If the recovery still fails, we fallback to the default label \textit{UnknownFunc}.

If the aforementioned process was completed successfully, the external call statement will be converted to \textit{UniswapV3.flash(...args)}.
And we can produce a textual summary for smart contracts containing control-flow and semantic call information simply by concatenating $F_i = (BB^i_0, \cdots, BB^i_{|F_i - 1|})$ $\in \mathcal{F}$ together in the form shown in Figure \ref{fig:ir_pscft_comparison}.

\subsection{Classifier Design}
\label{Classifier Design}

\subsubsection{Transformer}
\label{Semantic Transformer}

We train a transformer-based Natural Language Processing (NLP) model $\mathcal{M}^{transformer}$ solely on the PSCFT constructed from the contracts in our dataset to learn the semantics and behaviours of the implementation of both benign and adversarial smart contracts. Following the standard architecture of transformers~\cite{vaswani2017attention}, we design:

\begin{enumerate}
    \item A word2vec~\cite{mikolov2013efficient} embedding layer for positional encoding of PSCFT syntax and semantics,
    \item Multiple encoder layers chained together, each made up of a multi-head attention layer followed by a feed-forward layer,
    \item Pooling and dense layers for outputting classification results.
\end{enumerate}
To ensure the model's generalization ability and improve training efficiency, we implement regularization techniques such as early-stopping and dropouts to prevent model over-fitting. Overall, this transformer model provides a semantic view of smart contracts.

\subsubsection{Ensemble Classifier}
\label{Ensemble Classifier}

Aiming to embed the features selected based on our empirical observations into the semantic view of the transformer model, we design an ensemble classifier that merges the classifier models trained on empirical features with the transformer model trained on PSCFT together to produce a meta classifier model with better predictive performance.

\textbf{Candidate Classifiers.} Once all features have been extracted, we can proceed to train a supervised learning classifier using a variety of ML algorithms, outputting multiple candidate models for further evaluation. These models $\mathcal{M} = \{M^{classifier}_{LR}, M^{classifier}_{DT}, M^{classifier}_{RF}, M^{classifier}_{XGBM}\}$ include classical methods such as Logistic Regression and Decision Trees, ensemble learning methods such as bagging-based Random Forest and XGBoost which is based on boosting.

Due to the limited quantity of adversarial samples, to mitigate the bias introduced by class imbalance, we employ the ADASYN \emph{oversampling} method~\cite{he2008adasyn} that synthesizes additional samples to bring the number of adversarial samples closer to that of the benign ones in the training dataset.

For the fund source feature, which consists of categorical data (see \S\ref{DeploymentFeatures}) with low cardinality, we apply one-hot encoding. To facilitate better comparability among input values in numerical terms of varying dimensions, we standardize the dataset using Z-score normalization. This process brings each feature value around to the centre of $0$ with a standard deviation of $1$.

After training has completed for each of the candidate classifiers, we evaluate their performance using a function $f: \mathcal{M} \rightarrow \mathbb{R}$ that incorporates metrics such as Recall and F1-Score to determine an optimal model $M^{classifier}$ such that $M^{classifier} = \text{argmax}_{M \in \mathcal{M}}f(M)$ for the meta classifier training and inference in the next stage.
 
\textbf{Meta Classifier.}
\label{Meta Classifiers.} Using the \emph{stacked generalization (stacking)} machine learning algorithm, we design a \emph{blender} that stacks $M^{transformer}$ and $M^{classifier}$ together to form a higher level model $M^{meta}$ known as a meta classifier. By feeding the predictions produced by the two base models into the meta model, we are able to 1) Combine the strengths of each model and automatically learn the optimal way to construct the inference result based on model significance, 2) Optimize the final classification boundaries to achieve more accurate detection of adversarial contracts.

Given that the meta classifier operates on low-dimensional inputs (only the two probability values between $0$-$1$ outputted from the $M^{classifier}$ and $M^{transformer}$, indicating the likelihood of a contract being adversarial) and without incorporating any additional information about the contracts, employing a simple model is both sufficient and effective. This not only ensures efficiency but also mitigates the risk of overfitting by limiting model complexity. Accordingly, we only consider 
simple models, including K-Nearest Neighbors (KNN), Logistic Regression (LR), Support Vector Machines (SVM), and Decision Trees (DT), and evaluate their performance to select the most suitable meta classifier in experiments.
\section{Evaluation}
\label{evaluation}

We implement \name~in Python using open source machine learning frameworks PyTorch and Tensorflow to achieve both CPU and GPU-accelerated training and inference of our classifier models. All experiments are conducted on a server running a 3.70GHz Intel(R) Xeon(R) processor with 32 threads and a NVIDIA A100 GPU with 40GB of VRAM.

\subsection{Experiment Setup}

\subsubsection{Datasets}
\label{datasets}

Token and proxy contracts are two of the primary types of contracts used on blockchains, while they are commonly used by developers, most adversaries do not rely on those types of contracts for their core attack logic. To identify contract types, we perform a simple check based on PSCFT to determine whether a contract follows known standards such as ERC20 and ERC1967. By analyzing whether the contracts match common standards, we found that in the dataset we built, $184,386$ benign samples were token contracts, $10,037$ were proxy contracts, and none of the adversary contracts belonged to these two categories. Therefore, we remove them from both our training and testing dataset the same way as it was in previous work Forta~\cite{FortaNetwork}.

A distinguishing aspect of our research when compared to traditional supervised machine learning problems is the explicit chronological order of contract deployment. Contracts deployed over a different periods of time may exhibit varying characteristics and vulnerabilities, meaning that adversarial contracts may evolve and change over time as well. One of \name's objectives is to provide a ML classifier model that is pre-trained on past DeFi incidents but capable of detecting potential future attacks. Consequently, 
we sort the contract samples chronologically and use the earliest $80\%$ of the contracts for training and reserve the latest $20\%$ for testing. Out of the $80\%$ of contracts used as training data, we take the last $25\%$ of most recent samples as the training set for the meta classifier, and the rest for the transformer and candidate models.

\subsubsection{Research Questions}

\begin{table}[t]
\caption{Performance breakdown of different models.}
    \centering
    \small 
    \begin{tabularx}{\linewidth}{cXXXXXX} 
        \toprule
        \textbf{} & \textbf{Model} & \textbf{Accuracy} & \textbf{Precision} & \textbf{Recall} & \textbf{F1-Score} & \textbf{FPR} \\
        \midrule
        \multirow{4}{*}{\makecell{Candidate\\Classifiers}} 
        & LR & 0.9925 & 0.8788 & 0.7838 & 0.8286 & 0.0025 \\
        & DT & 0.9838 & 0.6222 & 0.7568 & 0.6829 & 0.0108 \\
        & RF & 0.9913 & 0.8594 & 0.7432 & 0.7971 & 0.0029 \\
        & \textbf{XGBoost} & \textbf{0.9932} & \textbf{0.8333} & \textbf{0.8784} & \textbf{0.8553} & \textbf{0.0041} \\
        \midrule
        \multirow{2}{*}{\makecell{NLP\\Classifier}} 
        & Forta & - & 0.8778 & 0.5520 & 0.6208 & - \\
        & \textbf{Transformer} & \textbf{0.9876} & \textbf{0.7429} & \textbf{0.7027} & \textbf{0.7222} & \textbf{0.0057} \\
        \midrule
        \multirow{4}{*}{Meta Classifier} 
        & LR & 0.9950 & 0.9683 & 0.8133 & 0.8841 & 0.0006 \\
        & DT & 0.9913 & 0.7901 & 0.8533 & 0.8205 & 0.0054 \\
        & SVM & 0.9947 & 0.9143 & 0.8533 & 0.8828 & 0.0019 \\
        & \textbf{KNN} & \textbf{0.9953} & \textbf{0.9286} & \textbf{0.8667} & \textbf{0.8966} & \textbf{0.0016} \\
        \bottomrule
    \end{tabularx}
\label{tab:classifier_model_performance_comparison}
\end{table}

We attempt to address the following research questions:

{\begin{enumerate}
    \item \textbf{RQ1.} How effective is \name~in detecting the adversarial contracts used in \emph{DeFi attacks}?
    \item \textbf{RQ2.} How do different empirical features contribute to model decision-making and how does PSCFT improve performance compared to using raw IR?
    \item \textbf{RQ3.} How efficient can \name~be in identifying adversarial contracts?
    \item \textbf{RQ4.} How practical is \name~ in real-world scenarios?
\end{enumerate}

\subsection{RQ1: Effectiveness}
\label{effectiveness}

\subsubsection{Evaluation Metrics}
In addition to commonly used binary classification performance metrics such as \textit{Accuracy}, \textit{Precision}, \textit{Recall} and \textit{F1-Score}, we also calculate \textit{False Positive Rates} (FPR), which holds particular significance in our scenario.

\subsubsection{Evaluation Results}
\label{Experiment Results}

Table~\ref{tab:classifier_model_performance_comparison} presents our experimental results evaluating the performance of various ML models (\S\ref{Classifier Design}) for the \name~system, including four candidate classifiers, a transformer-based classifier, and four meta classifiers. Among the candidate classifiers, XGBoost achieves the highest F1-score ($0.8553$), while Random Forest has the lowest recall ($0.7432$). Overall, the XGBoost-based classifier outperforms other candidate classifiers, demonstrating the best performance across most metrics compared to others.
It achieves a false positive rate as low as $0.0041$ while maintaining a high effective detection rate for adversarial samples.
Other models also perform well. Except the one based on DT, the F1-scores of all other classifiers exceed $0.79$, proving the validity of our method. 
Building upon the results of the candidate models, different meta classifiers were constructed by combining the optimal candidate model XGBoost with the transformer. The results show that three of the four meta classifiers achieved higher F1-scores compared to the standalone XGBoost classifier. Among them, the KNN-based classifier demonstrated the best performance, achieving the highest F1-score ($0.8966$) while maintaining a low FPR ($0.0016$).

\subsubsection{Comparison with Previous Works}

We compare our approach with two state-of-the-art baselines: Forta~\cite{FortaNetwork}, a NLP-based method for detecting all adversarial contracts, and BlockWatchDog~\cite{yang2024uncover}, a specialized work focused exclusively on detecting reentrancy attack contracts.

\textbf{Universality.} To evaluate the universality of the \name~system, we compare its performance against BlockWatchDog, which specifically targets reentrancy attack contracts. We test our meta classifier on all the reentrancy related adversarial contracts in our test dataset, we observe that \name~successfully identified $83.33\%$ of them, which is on par with the results presented by~\cite{yang2024uncover}. This observation along with the experiment results presented in \S\ref{Experiment Results} show that \name~not only generalizes well across diverse types of \textit{DeFi attacks} but also maintains competitive performance when applied to specific attack categories.

\textbf{Performance.} The only similar work that uses machine learning for adversarial contract detection publicly available so far is the Forta Network~\cite{FortaNetwork}. We use Forta as the baseline for comparing the performance of our ML models.

Forta is an NLP-based method that relies solely on EVM bytecode, while our method is based on a combination of multi-dimensional features and semantic tokenization (PSCFT) of smart contracts. We perform a direct comparison between their model and our transformer model first, then we measure how much benefits can our meta classifier approach provide over Forta's implementation. As the results in Table~\ref{tab:classifier_model_performance_comparison} indicate, our transformer alone can achieve much higher F1-Score than Forta, demonstrating the advantage of our PSCFT design. \name's meta classifier pushes the boundary even further by combining the capabilities of transformer model with the capabilities of other candidate classifiers to achieve an F1-Score of $0.8966$.
As described in \S\ref{External Data Retrieval}, contract verification status feature introduces some bias into our dataset, to provide a fair comparison, we also evaluate our models with $verified$ feature removed. The results show that even without this feature, our classifiers still significantly outperform the most recent model published by Forta (released on February 6, 2023), achieving an F1-Socre of $0.8696$, highlighting the effectiveness of our solution.

\subsubsection{Time Series Cross-Validation}

As outlined in~\ref{datasets}, our evaluation respects the chronological order of contract deployment, accounting for both the evolving nature of adversarial behaviours and \name's design to detect future attacks based on historical data. Table~\ref{tab:classifier_model_performance_comparison} shows the model performance results obtained by using the entire dataset with a 4:1 chronological split. We identify XGBoost as the optimal candidate classifier and KNN as the best-performing meta model. To validate the most appropriate candidate and meta models for our system and ensure its robustness under varying data conditions, we conduct evaluations using expanding window cross-validation. During this process, the dataset is chronologically divided into five equal splits. The training window starts with the first split and progressively expands by including additional splits, while the subsequent split serves as the test set in each iteration.

After cross-validation, we conclude that XGBoost achieves the highest average F1-score (0.8173), demonstrating stable performance across varying data splits. While it did not always outperform others in every split, its overall consistency reaffirms its suitability as the optimal candidate classifier for our system. For the meta classifiers, 
which combine the optimal candidate model XGBoost with the transformer, the KNN-based model achieves the highest average F1-score ($0.8355$). However, its advantage over other models were marginal, suggesting that the choice of meta classifier models has limited impact on overall performance. 

\subsection{RQ2: Interpretability}
\label{feature_analysis}

\subsubsection{Empirical Features}

Understanding the weight distribution of features is useful for gaining insights into the underlying decision-making process of the models. To this end, we apply the interpretability method SHAP (Shapley Additive exPlanations)~\cite{lundberg2017unified} to analyze the importance of empirical features for each candidate model.

The results reveal that both \textit{deployment} and \textit{implementation} features play significant roles in guiding the models to a decision. In the best-performing XGBoost model,
the top three most important features are \textit{avg\_token\_call\_count}, \textit{verified} and \textit{fund\_from\_Anonymous}, all of which also ranked within the top 10 across other candidate models. Additionally, \textit{balanceOf\_call\_count}, \textit{fund\_from\_Safe} and \textit{func\_count} consistently appear within the top 15 in at least three candidate models, indicating their stable contribution to the model's decision-making process.
The notable importance of these features may be attributed to their strong correlation with typical attack patterns, allowing the model to effectively capture behavioural characteristics of adversarial contracts.

\subsubsection{PSCFT}

The design of PSCFT abstracts the IR of a contract into a more concise and semantically meaningful format. To concretely demonstrate how this design enhances model performance, we compare the transformer classifier trained on PSCFT with one trained on the raw IR generated by Gigahorse. The results show a significant performance gap, where the model using raw IR achieves an F1-score of only 0.4068, substantially lower than the 0.7222 achieved by the PSCFT-based model.

This performance discrepancy highlights the limitations of directly using raw IR, which contains numerous low-level instructions (e.g., arithmetic operations, memory reads/writes) and excessive variables that obscure critical call patterns and their control flows. Such extraneous details increase input complexity, making it harder for the model to identify meaningful behavioural patterns relevant to adversarial contract. Our pruning process can effectively remove redundant instructions and enable semantic recovery, allowing PSCFT to focus on critical external calls and control flows, leading to clearer and more reliable model decisions and improved prediction accuracy.

\subsection{RQ3: Efficiency}
\label{efficiency}

\name~ aims to detect adversarial contracts and enable rapid responses within the rescue time frame (see \S\ref{Rescue Time Frame}). To this end, we evaluate the efficiency using all adversarial samples.

We employed a linear detection pipeline, which processes each sample through the following steps: 1) retrieve deployment features via external APIs, 2) execute our custom-designed Souffle Datalog and PSCFT generation scripts to extract implementation features, and 3) preprocess the data and feed it into pre-trained XGBoost and transformer classifiers. Their outputs are then combined by a pre-trained meta-classifier for the final prediction. We measure the time taken at each step.

The results showed that, on average, external API requests took $3.78$ seconds, while data preprocessing and model inference required only $0.01$ seconds. The most time-consuming step is binary lifting and analysis. Except for one large contract (taking $458$ seconds), most PSCFTs were generated under $20$ seconds, averaging $7.52$ seconds. Furthermore, by considering the \textit{rescue time frames} of the corresponding DeFi attacks, we found that in $92.8\%$ of the cases, our system successfully detected the adversarial contract before the adversarial transaction occurred. Notably, efficiency can be further improved by employing multi-threading and local blockchain nodes, which could substantially reduce the time required to obtain deployment features.

\subsection{RQ4: Practicality}

\noindent
\textbf{Setup}
We conducted live experiments targeting the \textit{Ethereum mainnet} across two time periods, 14–18 August 2024 and 23 January–9 February 2025, by deploying the \name~system to the server. The system continuously monitored the transactions in each new block via RPC providers~\cite{Alchemy} in near real-time. Upon detecting a contract creation (through a normal transaction), scripts were executed to extract features, which were then preprocessed and fed into the pre-trained classifiers (XGBoost and KNN, the best-performing setup as identified in RQ1). If classified as adversarial, the contract address was recorded for further verification.

\noindent
\textbf{Review Process}
We use a meticulous review process as follows to verify the outputs:

\begin{enumerate}
    \item Wait for $2$ weeks after each experimental period to ensure attacks have been confirmed;
    \item Manually inspect follow-up transactions involving the contracts to validate adversarial nature;
    \item Analyze and record key data, e.g. the financial damage caused by confirmed attacks, the time window between contract deployment and corresponding adversarial transaction.
\end{enumerate}

\begin{table}[t]
\caption{Representative true positive adversarial contracts detected during the live experiment, involved in attacks triggered by different root causes. \textit{Window} indicates the time interval between contract deployment and the adversarial transaction. \textit{Runtime} denotes the detection time, and \textit{Victim} specifies whether the victim address was exposed during contract creation.}
\centering
\small
\begin{tabularx}{\linewidth}{ccccccc}
\hline
\textbf{Address} & \textbf{Time}  & \textbf{Window} & \textbf{Runtime} & \textbf{Victim} & \textbf{Root Cause} & \textbf{Loss (\$)} \\
\hline
\href{https://etherscan.io/address/0x9f27901f9cc30bfe55b427f8c706f8b1d8b46d69}{0x9f2790} & 2024/8/16 & 25344s & 11.53s & No & Price Manipulation & 15.5k \\
\hline
\href{https://etherscan.io/address/0x90744c976f69c7d112e8fe85c750ace2a2c16f15}{0x90744c} & 2024/8/16 & 636s & 4.84s & Yes & Stale Oracle & 21.5k \\
\hline
\href{https://etherscan.io/address/0x1526195738e545807207227d49ca947fe021ec97}{0x152619} & 2024/8/18 & 336s & 5.67s & Yes & Access Control Flaw & 1.2k \\
\hline 
\href{https://etherscan.io/address/0x172133556107e05f281ce2e75b6a6fbe0965e055}{0x172133} & 2025/1/23 & 3504s & 6.12s & No & Absence of Input Validation & 2.5k \\
\hline
\href{https://etherscan.io/address/0x203f200fb833af6938e5994ae6e71d05734eff7c}{0x203f20} & 2025/2/6 & 120s & 14.35s & Yes & Arbitrary Transfer & 37k \\
\hline
\href{https://etherscan.io/address/0xbfe1025a910afa30741877aa97dabee86a6d1f06}{0xbfe102} & 2025/2/6 & 96s & 6.82s & Yes & Reentrancy & 8.7k \\
\hline 
\end{tabularx}
\label{tab:top_attacks}
\end{table}

\noindent
\textbf{Results}
During the experiment period, LookAhead analyzed a total of $23,463$ contracts and flagged $129$ as adversarial. Among these, $62$ involved follow-up transactions, in which we finally found $26$ true positives that were used to perform attacks, resulting in over \$150,000 USD in financial losses. These attacks were driven by various underlying issues, with Arbitrary Transfer~\cite{ma2023pied} being the most common, involving 6 contracts. Table~\ref{tab:top_attacks} presents representative adversarial contracts involved in attacks stemming from different root causes, highlighting the universality of our approach, which effectively identifies various types of adversarial behaviours. 

In terms of detection efficiency, except only two cases where the deployment and attack transaction were confirmed within the same block, \name~produced a response for all the attack events before the attacker actually initiated the adversarial transaction.
Additionally, we observed most of the false positives are related to arbitrage, showing that arbitrage behaviour resembles some patterns in adversarial activities, such as reliance on flashloans and frequent token-related function calls.
To evaluate potential false negatives, we reviewed the alerts posted by security company (e.g. \cite{blocksec, certik, slowmist}) during the experiment period, as manually checking for false negatives among a large number of contracts is challenging.
This process identified four reported attacks that fell within our detection scope (\S\ref{defi attacks}), all of which had already been correctly flagged.

The fact that, during our live experiment, \name~was able to respond in a timely manner before the majority of real-world attacks occurred, and successfully detected all externally reported attacks, further demonstrates its practicality in real-world scenarios.

\section{Discussion}
\label{discussion}

\subsection{Threats to Validity}
\label{threats to validity}
1) Limited by the APIs given by the blockchain explorers, we cannot determine whether contract verification occurred upon deployment;
2) Our dataset suffers from a class imbalance of approximately $1:40$ due to the scarcity of adversarial contracts, which may affect model performance even after oversampling;
3) Benign contracts were selected based on heuristic approach. There is still a possibility that we have mislabelled some adversarial contracts as benign ones;
4) As discussed in \S\ref{Benign Contract Collection}, our study does not consider contracts deployed via internal transactions. The inherent differences between the two deployment methods may affect certain \textit{deployment features}, introducing potential bias to our classifier when detecting contracts deployed via internal transactions.




\subsection{Limitations}

In addition to the inherent limitations of our threat model definition (\S\ref{thread model}), there are mainly two aspects of \name's design that could potentially hinder its performance.

\noindent
\textbf{Detection Evasion}
Since our classifier relies on extracted features and historical attack patterns, it remains susceptible to evasion techniques. In the cat-and-mouse game of DeFi attack and defense, attackers may adapt their strategies to bypass our feature-based and model-driven approach. For instance, they could adopt code obfuscation to alter the contract’s CFG or utilize KYC-required exchanges as funding sources, making it harder for our system to recognize adversarial intentions.

\noindent
\textbf{NLP-based Approach} 
\name's NLP-based models may not make full use of the control flow graph structure present in smart contract code. More advanced ML architectures such as Graph Neural Network (GNN) could be used in future works in combination with a more intricate representation design that capture both intra-function CFGs and inter-function call relationships.

\section{Related Work}
\label{related work}

\noindent
\textbf{Adversarial Transaction Detection.}
Prior research primarily focus on detecting adversarial transactions used in DeFi attacks. Some works~\cite{chen2020soda,grossman2017online,rodler2018sereum,wu2022time} modify EVM to obtain more detailed transaction execution information for detecting adversarial transactions. 
BlockGPT~\cite{gai2023blockchain} applies large language model to detect abnormal transactions.
Qin et al.~\cite{qin2023towards} introduce the Execution Property Graph to enable timely detection. However, these methods become ineffective in identifying attacks when attackers leverage private mempool services to initiate adversarial transactions.

\noindent
\textbf{Adversarial Contract Detection.} The study of adversarial contract detection in DeFi attacks remains relatively underexplored. Forta~\cite{FortaNetwork} adopts an NLP-based approach for adversarial contract detection; however, its reliance on simple bytecode-level analysis results in limited effectiveness. Beyond Forta, the only comparable work is BlockWatchDog~\cite{yang2024uncover}. However, their approach is restricted to reentrancy attacks and relies on prior knowledge of victim addresses, thereby limiting its practical utility in real-world scenarios.
Our method integrates multi-faceted features with semantic tokenization to allow more effective detection across various DeFi attack types.

\section{Conclusion}
\label{conclusion}

In this paper, we propose a new framework for effectively detecting DeFi attacks. We build the first large-scale comprehensive dataset for training classifiers for the binary classification of adversarial contracts.
Through the analysis of DeFi attack patterns, we extract a multi-dimensional feature set along with a special tokenization technique that is able to capture the intrinsic behaviours hidden in adversarial contracts. Utilizing these features, 
we carefully design multiple classifiers and evaluate them comprehensively. Experiments show that our method performs exceptionally well in detecting adversarial contracts compared to the previous state-of-the-arts with our models reaching F1-Scores as high as $0.8966$ and a false positive rate as low as $0.16\%$ on our labeled datasets.

\section{Data Availability}

To promote open science, we have made our dataset and source code publicly available~\cite{lookahead}.

\begin{acks}
This work is sponsored in part by National Key Research and Development Program of China (2023YFB2704000).
\end{acks}

\bibliographystyle{ACM-Reference-Format}
\bibliography{refs}


\end{document}